\documentclass[12pt]{article}
\usepackage{amsmath,amsfonts,amssymb,amsthm,newlfont,exscale,color}
\usepackage[dvips]{graphicx}

\newcommand{\mb}{\bar{\mu}'(0)}
\newcommand{\myu}[1]{\bar{\mu}(#1)}
\newcommand{\lam}[1]{\bar{\lambda}(#1)}
\newcommand{\ro}[1]{\bar{\rho}(#1)}
\newcommand{\taau}[1]{\bar{\tau}(#1)}

\begin{document}
\bibliographystyle{unsrt}
\title{Extinction times for birth-death processes: exact results,
continuum asymptotics, and the failure of the Fokker-Planck approximation}

\maketitle

\begin{center}

Charles R. Doering, Khachik V. Sargsyan \\
{\it Department of Mathematics and Michigan Center for Theoretical
Physics \\ University of Michigan, Ann Arbor, MI 48109-1109 USA} \\

\medskip
and \\
\medskip

Leonard M. Sander \\
{\it Department of Physics and Michigan Center for Theoretical Physics \\
University of Michigan, Ann Arbor, MI 48109-1120 USA} \\

\end{center}

\bigskip

\begin{abstract}

We  consider extinction times for a class of birth-death processes commonly
found in applications, where there is a control parameter which
determines whether the population quickly becomes extinct, or rather
persists for a long time.  We give an exact expression  for the discrete
case and its asymptotic expansion for large values of the population. We
have results below the threshold, at the threshold, and above the
threshold (where there is a quasi-stationary state and the extinction time
is very long.) We show that the Fokker-Planck approximation is valid
only quite near the threshold. We compare our analytical results to
numerical simulations for the SIS epidemic model,  which is in the class
that we treat. This is an interesting example of the delicate
relationship between discrete and continuum treatments of the same
problem.

\end{abstract}

\newpage


\section{Introduction}

Birth-death processes are widely used as a description of processes in
physics, chemistry \cite{Gardiner83,Kampen92}, population biology
\cite{Nasell01}, and many other areas. These are Markov processes defined on
states which we label by $n=0,1,...,R$ where $R$ denotes the largest
value allowed (which could be $\infty$). They are defined by the birth
and death rates:
\begin{eqnarray}
\lambda_n & &  \quad n\to n+1, \nonumber \\
\mu_n & & \quad n \to n-1.
\label{probs}
\end{eqnarray}

The processes we will consider have an absorbing state (`extinction')
which we put at $n=0$; That is, $\lambda_0=0$. In what follows we will
be mainly concerned with the mean time to extinction, $\tau_k$, i.e.,
the mean first passage time to the state $n=0$ starting at $n=k$. For
any Markov process with an absorbing state, extinction will occur as $t
\to \infty$ with unit probability. For example, if $n$ denotes the size
of a population of organisms, we seek the mean time to biological
extinction.

In this paper we will give exact expressions for the extinction time
for a class of birth-death processes, and asymptotic expressions for
cases where the typical $n$ is large, i.e., in a `continuum' limit.
We will investigate the validity of a popular approximation, the 
Fokker-Planck or diffusion method \cite{Gardiner83}. We will see that the 
Fokker-Planck method gives the correct asymptotic 
continuum behavior of $\tau$ only in very special circumstances.

To fix ideas, consider the following two processes taken from the
literature of  epidemiology and population biology.
\begin{itemize}
\item The Susceptible-Infected-Susceptible (SIS) model of epidemiology 
\cite{Jacquez93}: Imagine 
a population of size $N$ within which $n$ individuals suffer from an 
infection, and the rest, $N-n$, are susceptible. 
Suppose the infection rate per contact is $\Lambda/N$, the number of 
contacts is $n(N-n)$ and that the recovery rate is unity (fixing the unit 
of time).
A recovered individual immediately becomes susceptible. 
Then:
\begin{eqnarray}
\lambda_n &=& \Lambda n(1-n/N), \nonumber \\
\mu_n &=& n.
\label{SISeqn}
\end{eqnarray}
At the deterministic (non-stochastic, continuum) level then there may be a 
non-zero steady state number, $n_e$, of infected individuals, the solution 
of $\lambda_n=\mu_n$.  
In this SIS model $n_e = N(1-1/\Lambda)$, provided $\Lambda > 1$.
This model has a threshold, $\Lambda=1$, above which the infection
persists in the contimuum approximation.
When the $\Lambda \le 1$ the infection dies out. 
In the stochastic model, however, above threshold the number of infected 
individuals remains near $n_e$ for a long time (the quasi-stationary 
state) before eventually going extinct \cite{Nasell99}.

\item A logistic model from ecology \cite{Grasman97}, often called the Verhulst model:
This population
dynamics model assumes a birth rate per individual, $B$, and unit death
rate, per individual. 
In order account for competition for resources, the death rate is assumed to increase 
proportional to $n^2$. 
We write:
\begin{eqnarray}
\lambda_n &=& Bn, \nonumber \\
\mu_n &=& n + (B-1)n^2/N,
\label{Verhulst}
\end{eqnarray}
defining the carrying capacity, $N$.
At the deterministic level, $n_e=N$  provided $B>1$. 
In the continuum, for $B>1$ the population stabilizes at $n_e$ while for 
$B\le 1$ it goes extinct. 
In the stochastic model there is a quasi-stationary state for $B>1$ in 
which the population fluctuates near $n_e$ before eventually going 
extinct.
\end{itemize}

These two examples are representative of the class of models that we
now consider.
In both examples there is a large number, $N$, and we assume that both 
$\lambda$ and $\mu$ involve such a number in a special way:
\begin{eqnarray}
\lambda_n &=& N\overline\lambda(x), \nonumber \\
\mu_n &=& N\overline\mu(x),
\label{scale}
\end{eqnarray}
where $x = n/N$ and $\overline\lambda, \overline\mu$ are smooth functions 
of $x$. 

These processes have the following properties: First, we assume that
$\lambda_n = N\overline\lambda(x)$ is concave downward (or linear) and
$\mu_n = N\overline\mu(x)$ concave upward (or linear).
(We will not consider the general degenerate case where both functions are 
linear).
Both functions are taken to have finite non-zero slopes near $n=0$.
The processes are most interesting when there is a control parameter so
that there can be an intersection of the two curves (super-threshold), or
not (sub-threshold) depending on the parameter.
We are also interested in the case when the parameter is very near 
threshold, in a sense that we will define below; see Figure 1.

\begin{figure}
\parbox{2in}{\includegraphics[width=2in,height=2in]{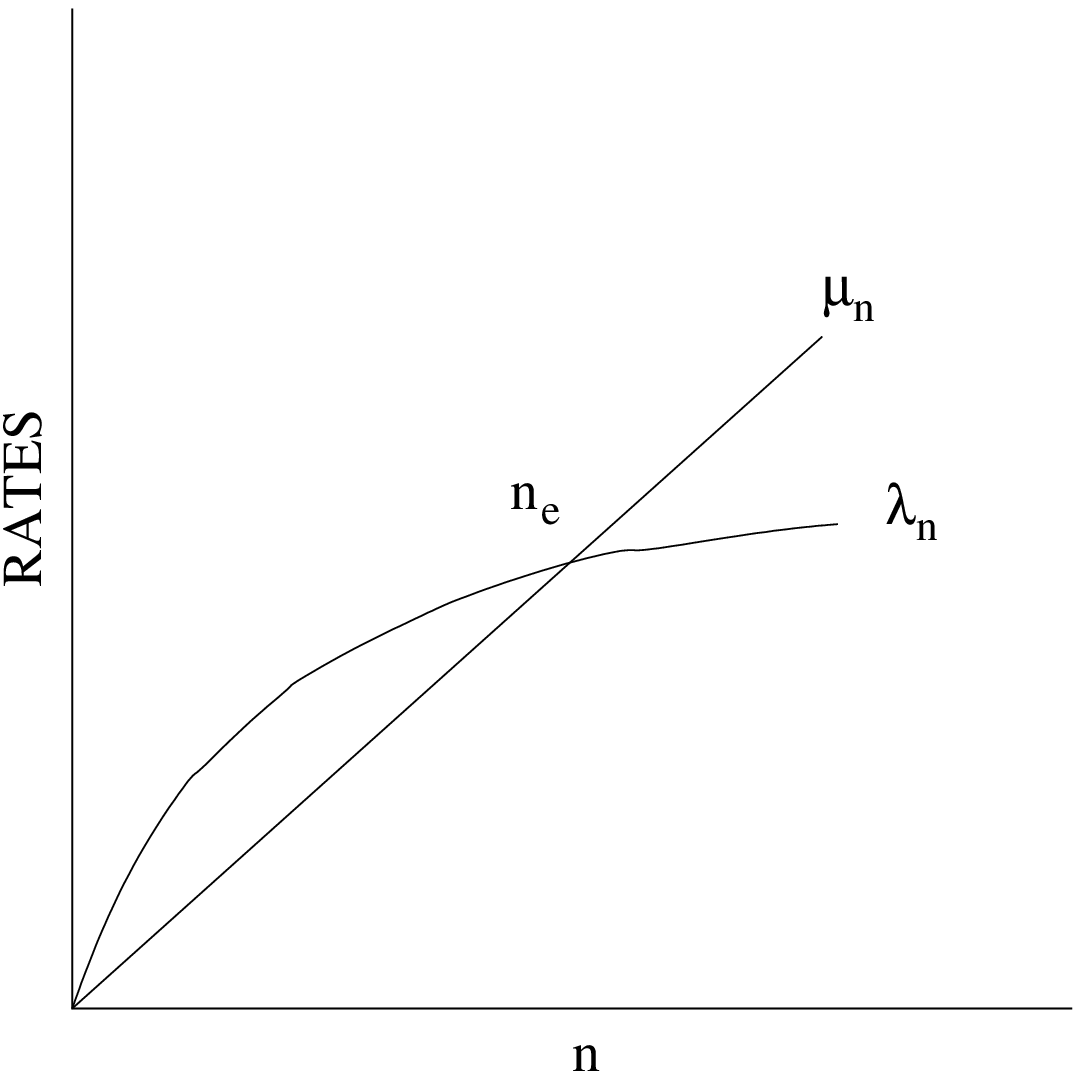} }
\hspace{.1in}
\parbox{2in}{\includegraphics[width=2in]{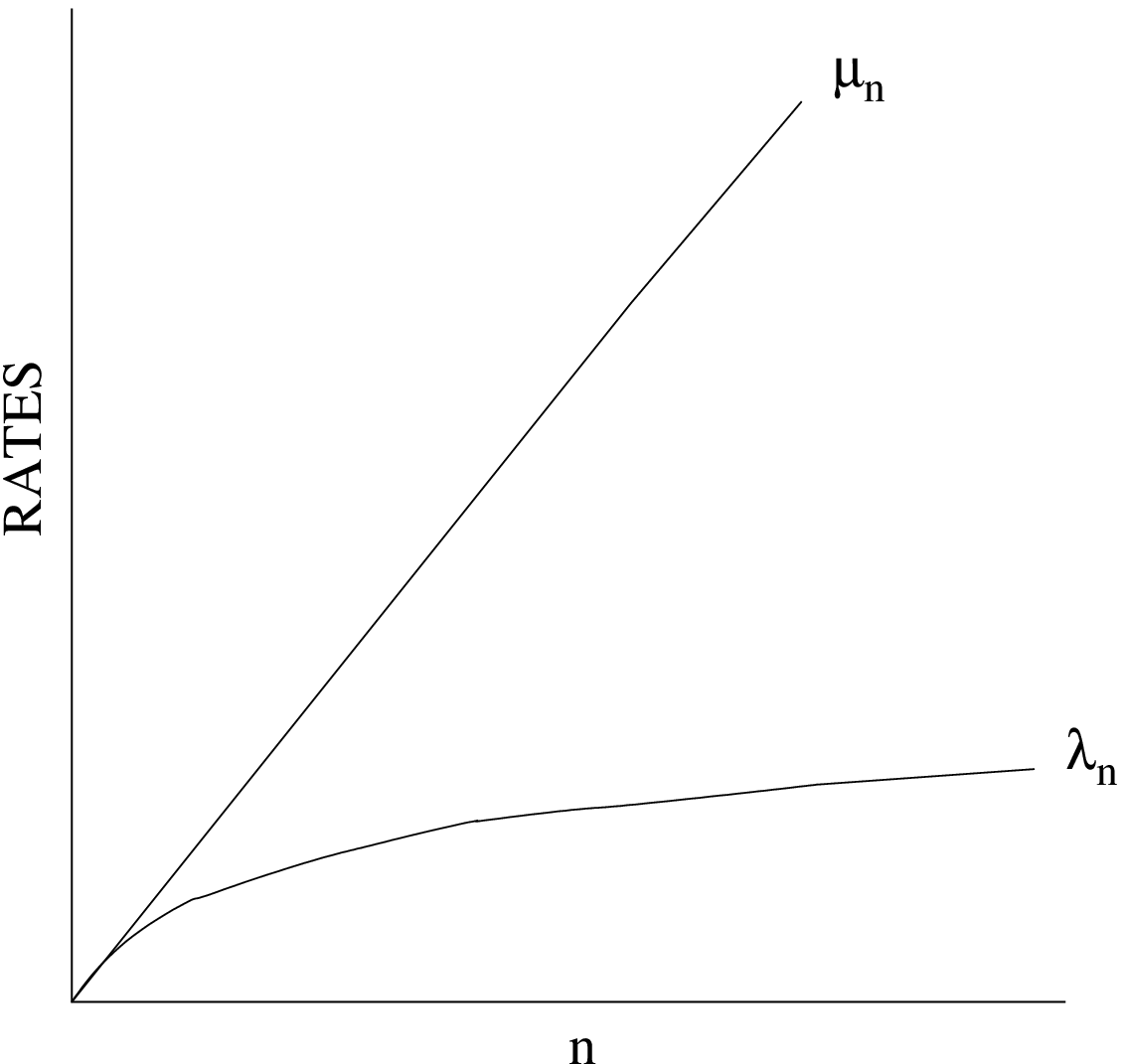}}
\hspace{.1in}
\parbox{2in}{\includegraphics[width=2in]{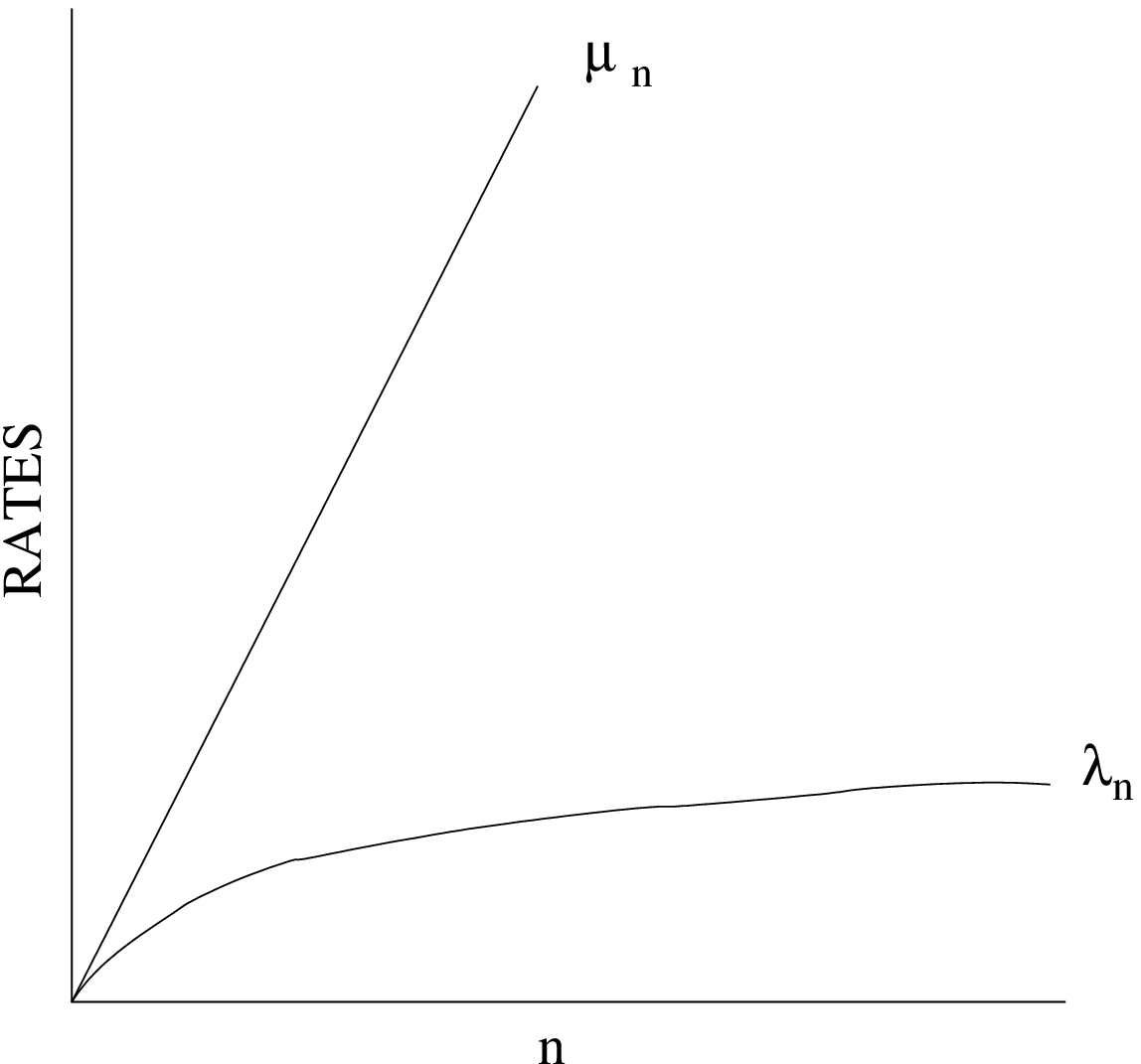}}
\caption{Regimes for the rates $\lambda_n, \mu_n$.
a.) Above threshold. b.) Near threshold c.) Below threshold.}
\end{figure}

We are interested in the mean time to extinction starting at state $n$. 
The probabilities, $\pi_n(t)$, for the various states obey the master equation:
\begin{equation}
d\pi_n(t)/dt = \lambda_{n-1}\pi_{n-1}(t) -(\lambda_{n}+\mu_n)\pi_n(t)
+\mu_{n+1}\pi_{n+1}(t).
\label{master}
\end{equation}
It is an elementary exercise \cite{Gardiner83,Kampen92} to show that 
$\tau_n$ obeys:
\begin{equation}
-1 = \lambda_{n}\tau_{n+1} - (\lambda_{n}+\mu_n)\tau_n
+\mu_{n}\tau_{n-1}
\label{taueq}
\end{equation}
We will solve this equation exactly, below, and give expressions for the
asymptotic behavior of $\tau_k$ as $N \to \infty$. In the context of the
SIS model there is a very large literature on this question
\cite{Weiss71,Oppenheim77,Nasell99,Andersson98}. Our general expression agrees
with the main results of Ref. \cite{Andersson98} for this case. However, in 
some cases we find different results, see below.

For $N>>1$, near the continuum limit, it is tempting to work directly with 
Eq.
(\ref{taueq}), and try to replace it by a differential equation in $x$.
A naive Taylor expansion gives:
\begin{equation}
-1 \ = \ f(x)T'(x) + \frac{1}{2N}g^2(x)T''(x)
\label{FPE}
\end{equation}
where $x=n/N$ is the continuum variable, $T(n/N) = \tau_n$, and we define:
\begin{eqnarray}
f(x)&=& \lam{x}-\myu{x} \nonumber \\
g^2(x) &=&\lam{x}+\myu{x} 
\label{FPtau}
\end{eqnarray}
The smooth functions $f(x)$ and $g^2(x)/2N$ are, respectively, the 
drift (sometimes called the `force') and the diffusion coefficient.

The operator on the right-hand side of the first equation above is the 
adjoint of the operator in the Fokker-Planck equation (FPE):
\cite{Gardiner83,Kampen92}:
\begin{equation}
\partial_t P(x,t) = (1/2N)\partial_{xx} (g^2 P) - \partial_x(fP)
\label{FPEpde}
\end{equation}
This is the evolution equation for the probability transition density of a
random walker with diffusion coefficient $g^2/2N$ subject to a drift 
velocity $f$. 
This approach, which is particularly popular in the natural sciences, is 
based on the observation that we can think of the birth-death process as a 
biased random walk in $n$. 
From the work of Einstein and Schmoluchowski, we can describe a continuum 
random walk with a diffusion equation, namely (\ref{FPEpde}). 
The extinction time is the first passage time to the origin of the random
walker.  This approach is attractive because the ordinary differential equation in 
(\ref{FPtau}) is very easy to solve, and gives a tractable formula for $T$ 
in the limit of large $N$. 
Grasman and collaborators used this approach for the logistic model
\cite{Grasman97,Grasman98}, and apparently verified their results by numerical calculations.
It is also possible to convert Eq. (\ref{FPEpde}) into a stochastic differential equation of the Langevin type \cite{Kampen92}, which has attractive numerical properties.

However, as we will see below, this series of manipulations only gives
the correct asymptotic behavior of $\tau$ \emph{under very special
circumstances}, namely when the two terms in Eq. (\ref{FPE}) are
comparable in size. In our case, this means that $f = \lambda - \mu$ is
very small. There are several ways in which this can happen. For
example, if we are interested in small fluctuations near $x_e \equiv
n_e/N$, then we can almost certainly use a Gaussian approximation and
get reliable results (though we do not verify this explicitly here).
This is what Van Kampen \cite{Kampen92} calls the `system size expansion', and it
is a standard method in applications.

For the extinction problem we expect that the important $n$'s
will be all those between $n=n_e$ and $n=0$. 
For the force to be small for the whole range we must be near threshold, 
as in Figure (1b).  As we will see below, 
in order for Eq. (\ref{FPE}) to correctly describe the asymptotic behavior we need
$\rho_n = \lambda_n/\mu_n = 1+ {\cal O}(1/N^{1/3+\epsilon}), \epsilon>0$, 
or, equivalently, $x_e = {\cal O}(N^{-1/3-\epsilon})$. 
We will discuss the work in Ref. \cite{Grasman97} below, and show that the case 
they treated was, in fact, near threshold. This accounts for the the agreement between their FPE results and numerical calculations.

This result is counterintuitive, and demonstrates the delicacy of the continuum limit even for the very simple class of processes that we consider here. Explicitly,  the continuum FPE is useful 
not for large populations, as one might guess, but only near threshold 
where the quasi-equilibrium population is \emph{much smaller} than $N$. 
The upshot is that a diffusion treatment is valid only when the drift is 
small. In fact, we will go further below: we will show that no diffusion 
description is possible for these processes (except just above threshold) in 
the sense that no diffusion equation can simultaneously give $\tau$ 
correctly and also give a good approximation to the quasi-steady state.

In the next section we will state our results. In the next section we give
some numerical illustrations for the SIS process.
In the final section we will discuss the implications of our findings for 
applications. 
We will relegate the actual computations to the Appendices.


\section{Results}

In this section we summarize our main results.  
First we briefly discuss the exact solution for the mean extinction time 
as a function of the initial $n$.
Subsequently we present the large $N$ asymptotic expansions of the 
solutions in the superthreshold, threshold and subthreshold cases.
\subsection{Extinction time}
The first problem is to solve the second order difference equation 
(\ref{taueq}) for $1 \le n \le R-1$ with absorption at site $n=0$ and a 
reflecting boundary condition ($\lambda_{R} = 0$) at site $n=R$, i.e.,
\begin{equation}
\tau_0=0, \ \ \ \ \tau_{R}-\tau_{R-1}=\frac{1}{\mu_R}.
\end{equation}
This is straightforward: let $\alpha_k=\tau_k-\tau_{k-1}$ for  
$k=1\ldots{R}$ so that
\begin{equation}
-1=\lambda_n\alpha_{n+1}-\mu_n\alpha_n
\end{equation}
for $1 \le n \le R-1$ with the single boundary condition 
$\alpha_R=\frac{1}{\mu_R}$.
Then the solution of this first order difference equation for $\alpha_m$, 
$1 \le m \le R-1$, is easily written down:  
\begin{equation}
\alpha_m = \frac{1}{\mu_m} + \sum_{j=1}^{R-m} \frac{1}{\mu_{m+j}} 
\prod_{i=1}^{j}\rho_{m+i-1}
\label{alpha}
\end{equation}
where
\begin{equation}
\rho_i = \lambda_i/\mu_i.
\end{equation}
Then the mean extinction time, $\tau_n$, is recovered from
\begin{equation} 
\tau_n = \sum_{m=1}^{n} \alpha_m.
\end{equation}
By regrouping the product in (\ref{alpha}) we find the  explicit 
solution with no approximation:
\begin{equation}
\tau_n = \sum_{m=1}^{n} 
\big[ 
\frac{1}{\mu_m} +  { \prod_{i=1}^{m-1} \frac{1}{\rho_i}  
\ \sum_{j=m+1}^{R} \frac{1}{\mu_j} \ \prod_{k=1}^{j-1} \rho_k  } 
\big].
\label{exact}
\end{equation}

\subsection{Expansion for large $N$}
Now write $\lambda_n=N\lam{n/N}$ and $\mu_n=N\myu{n/N}$ with 
$\bar{\lambda}(x)$ and $\bar{\mu}(x)$ uniformly smooth functions
on $[0, r]$ where $r=R/N$. 
We can define $\rho_n=\ro{n/N}$, where $\ro{x}$ is a bounded, smooth, 
and non-negative function on $[0, r]$. We will call the following quantity the `effective potential'
\begin{equation}
\Phi(x)=-\int_{0}^{x} \log{\ro{\xi}} d\xi.
\label{Phi}
\end{equation}
to facilitate comparison to the FPE, below.

For large $N$ the products in (\ref{exact}) can be estimated by the 
trapezoid rule of numerical analysis.
Using the continuous variables $z = j/N$ and $y = m/N$,
\begin{eqnarray}
\prod_{k=1}^{j-1}\rho_k 
&=& \exp{\left( \sum_{k=1}^{j-1}\log{\ro{k/N}} \right)} \nonumber \\
&=& \exp{\left( N\int_{0}^{z} \log{\ro{w}} dw 
- \frac{1}{2}(\log{\ro{0}} + \log{\ro{z}}) + {\cal O}(1/N) \right)} 
\nonumber \\
&=& \frac{1}{\sqrt{\ro{0}\ro{z}}} 
\exp{\left( N\int_{0}^{z}\log{\ro{w}} dw \right)} 
\times (1+{\cal O}(1/N)) \nonumber \\
&=& \frac{1}{\sqrt{\ro{0}\ro{z}}} \ e^{-N\Phi(z)}
\times(1+{\cal O}(1/N)).
\label{trap}
\end{eqnarray}
and similarly,
\begin{equation}
\prod_{i=1}^{m-1}\frac{1}{\rho_i} 
= \sqrt{\ro{0}\ro{y}} \ e^{N\Phi(y)} \times (1+{\cal O}(1/N)) 
\end{equation}
In these estimates and others following, the coefficients of the 
${\cal O}(N^{-\beta})$ error terms depend on the regularity of the rate 
functions $\lam{x}$ and $\myu{x}$.  We proceed under the assumption of 
sufficient smoothness so that  the estimates are valid.

In the following we write $x=n/N$ for the initial point.  
The large $N$ asymptotic behavior of $\taau{x} = \tau_n$ as given in 
(\ref{exact}) is very different for the superthreshold, threshold, and 
subthreshold cases:

\begin{itemize}
\item{\textbf{Superthreshold case}}

When $\bar\lambda{'}(0) >\bar \mu{'}(0)$ there is a unique `equilibrium' state $n_e$ where
$\lambda_{n_e} = \mu_{n_e}$, or equivalently a unique `deterministic
steady state' $x_e = n_e/N > 0$ where $\lam{x_e}=\myu{x_e}$.
(In the SIS and logistic models this corresponds, respectively, to the
conditions $\Lambda > 1$ and $B > 1$.) For the stochastic processes, 
the extinction time is exponentially large as $N \rightarrow \infty$. We find that
$\taau{x} \sim \tau_{n_e} = \taau{x_e} $ for $n={\cal O}(N)$, i.e., $x={\cal O}(1)$. 
Further:
\begin{equation}
\taau{x_e} 
= \sqrt{\frac{2 \pi \lam{0}\myu{0}}{N [\lam{x_e} \overline{\mu}'(x_e) 
- \overline{\lambda}'(x_e) \myu{x_e}]}} \times 
\frac{e^{-N\Phi(x_e)}}{\lam{0}-\myu{0} } 
 \times 
\left( 1+{\cal O}(\frac{1}{N}) \right)
\label{super}
\end{equation}

In this region $\taau{x}$ is independent of $x$.
For $x = {\cal O(1/N)}$ there is a boundary
layer where $\taau{x}$ depends on $x$. However, 
we have found a simple 
correction factor which allows us to give, for all $x \in 
[0,r=R/N)$, a uniform asymptotic approximation,
\begin{equation}
\taau{x} = \left( 1 - e^{-N \log{\ro{0}} x} \right) \taau{x_e}.
\label{blayer}
\end{equation}

\item{\textbf{Threshold case}}

Here $\lambda_n < \mu_n$ for $n \ge 1$, but the derivatives of $\lam{x}$ 
and $\myu{x}$ at 0 are equal.
(In the SIS and logistic models this corresponds,respectively, to 
$\Lambda=1$ and $B=1$.)
In this critical situation the dominant term in the large $N$ asymptotic 
expansion of the extinction time is $\propto N^{\frac{1}{2}}$:
\begin{equation}
\taau{x} = C \sqrt{\frac{\pi \ro{0}}{2 \Phi^{''}(0) 
\bar{\lambda}^{'}(0) \bar{\mu}^{'}(0)}} \times \sqrt{N} \ + \
\frac{\log{Nx}}{\bar{\mu}^{'}(0)} \ + \ {\cal O}(1)
\label{thres}
\end{equation}
for $n/N = x = {\cal O}(1)$. 
The  constant prefactor, $C \approx 1.57$, is the sum
\begin{equation}
\sum_{k=0}^{\infty}\frac{(2k)!}{(2^k k!)^2(2k+1)} = 1.5687\dots . 
\end{equation}

\item{\textbf{Subthreshold case}}

Below threshold, $\lambda_n < \mu_n$ for all $n \ge 1$, so $\ro{x} < 1$ 
for all $x \ge 0$.
Moreover, the derivative of $\lam{x}$ is strictly smaller than the
derivative of $\myu{x}$ for all $x \ge 0$.
(In SIS and logistic models this corresponds, respectively, to
$\Lambda < 1$ and $B < 1$.) The deterministic
version  go to extinction rapidly in this case. For the stochastic processes,
the extinction time is logarithmic in $N$ as $N \rightarrow \infty$:
\begin{equation}
\taau{x} = \frac{1}{\mb(1-\ro{0})}\log{Nx} + {\cal O}(1)
\label{sub}
\end{equation}
for $n/N = x = {\cal O}(1)$.
\end{itemize}


\section{Numerical estimates and the failure of the Fokker-Planck approximation}
In this section we compare our results to numerical calculations, and to the FPE approximation for the SIS model.
Our numerical results are based on a direct evaluation of the exact formula Eq. (\ref{exact}).
\subsection{Numerical and analytical results for the SIS model:}
As we pointed out above, an example of the family of models which concerns  us here is the
SIS model defined by:
\begin{equation}
\lambda_n = \Lambda n\big(1-\frac{n}{N}\big),
\qquad \mu_n=n.
\end{equation}
For this case:
\begin{equation}
\lam{x}=\Lambda x(1-x), \qquad \myu{x}=x,
\end{equation}
so that
\begin{equation}
\ro{x}=\Lambda(1-x)
\end{equation}
and
\begin{equation}
\Phi(x)=-\int_0^x
\log{\Lambda(1-\xi)}d\xi = (1-x) \log{\Lambda(1-x)} + 
x - \log{\Lambda}.
\end{equation}
It is easy to check that $\Phi(x)$ is a convex function.

The analytical results for this model based on Eqs. (\ref{super}), (\ref{thres}), and (\ref{sub}) 
are presented in the following table.
\[
\begin{tabular}{ccccc}

\qquad & $\Lambda$ & $\Phi(x)$ & \qquad $\tau_n$ for $n/N=x={\cal O}(1)$ & 
\\
  \hline
   \hline
\vspace{.1in}   
 superthreshold & $\Lambda>1$ & $\Phi^{'}(0)<0$ &
$\frac{\Lambda}{(\Lambda-1)^2}\sqrt{\frac{2\pi}{N}}
e^{N(\log{\Lambda}-1+1/\Lambda)} \times \left( 1+{\cal O}(1/N) \right)$ & 
Fig.\ref{superfig}\\
\vspace{.1in}

  threshold & $\Lambda=1$& $\Phi^{'}(0)=0$ &
  $1.5687\sqrt{\frac{\pi}{2}}
\sqrt{N}+\log{Nx} + {\cal O}(1)$ & Fig. \ref{thresfig}\\
\vspace{.1in}

     subthreshold & $\Lambda<1$ & $\Phi^{'}(0)>0$  &
$\frac{1}{1-\Lambda}\log{Nx} + {\cal O}(1)$ & Fig.\ref{subfig}\\

\end{tabular}
\]

\begin{figure}
\centerline{\includegraphics[width=4in]{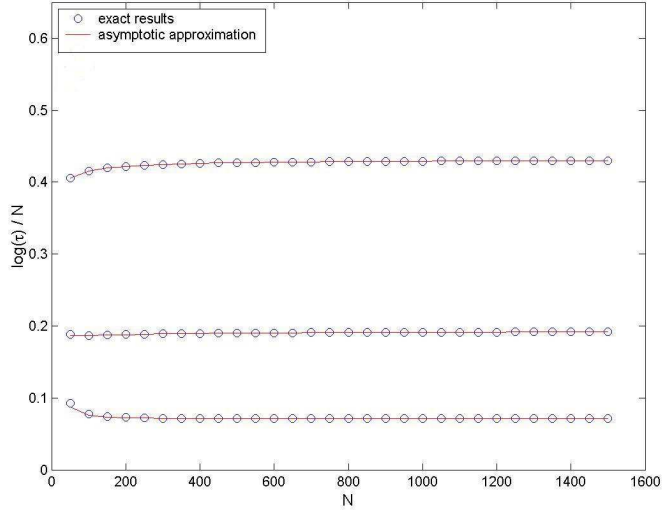}}
\caption{Comparison of analytical and numerical results for the SIS model in the superthreshold 
case. $\log\taau{x_e}/N$ is plotted as a function of $N$ for  $\Lambda= 1.5, 2, 3$ bottom to top.}
\label{superfig}
\end{figure}

\begin{figure}
\centerline{\includegraphics[width=4in]{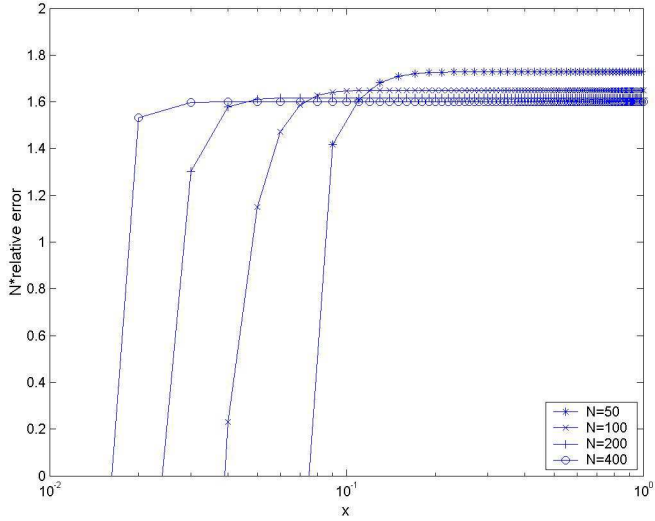}}
\caption{$N$ times the relative error for the superthreshold case of the SIS model as a function 
of $x$ for various $N$ and $\Lambda=3$.}
\label{error}
\end{figure}

The superthreshold case is compared with numerical simulations in Fig. (\ref{superfig}). Our formula agrees with the results of reference \cite{Andersson98} and with the exact results.

We can go further and test the validity of our error estimate in the first line of the table above, and also our treatment of the boundary layer. We define the relative error as $\bar \tau/\bar\tau_{asy} -1$, where  $\bar\tau_{asy}$ is given in Eqs. (\ref{super}), (\ref{blayer}). We plot $N$ times this quantity in Fig. (\ref{error}) to show that the relative error is of order $1/N$ and is uniform in $x$.

\begin{figure}
\centerline{\includegraphics[width=4in]{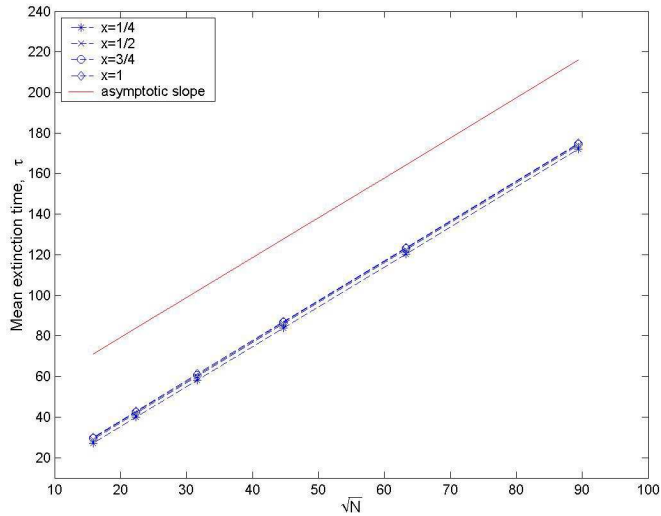}}
\caption{Comparison of analytical and numerical results for the SIS model in the threshold case. 
$\taau{x}$ is plotted as a function of $N^{1/2}$ for various values of $x$. The symbols are exact
results of Eq. (\protect\ref{exact}) and the dashed lines are guides for the eye. The solid line is 
the prediction for the coefficient of $N^{1/2}$ in Eq. (\protect\ref{thres}). The vertical position 
of the solid line is arbitrary. }
\label{thresfig}
\end{figure}

\begin{figure}
\centerline{\includegraphics[width=4in,height=3in]{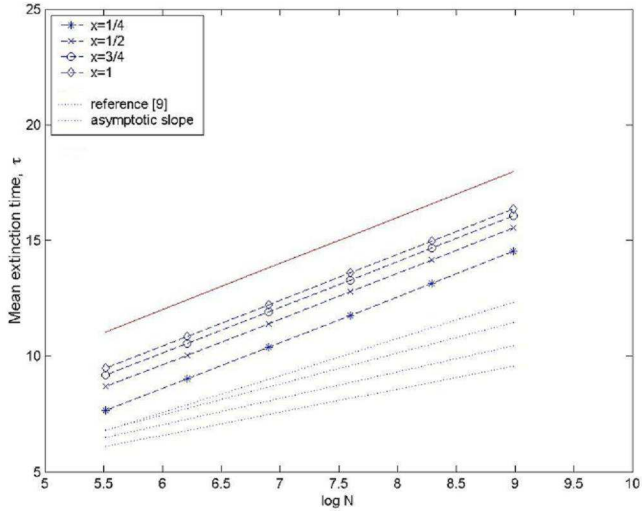}}
\caption{Comparison of analytical and numerical results for $\taau{x}$ in the subthreshold 
case of the SIS model for $\Lambda=1/2$. $\taau{x}$ is plotted as a function of $\log{N}$ for 
several values of $x$. The symbols are exact results and the dashed lines are guides for the eye. 
The solid line is the prediction of Eq. (\protect\ref{sub}). The vertical position of the solid 
line is arbitrary.  The results of \protect\cite{Andersson98} are also shown as dotted lines 
which correspond to $x=1/4, 1/2, 3/4, 1$ from top to bottom. }
\label{subfig}
\end{figure}

The threshold case is shown in Fig. (\ref{thresfig}); we find good agreement between our asymptotic formula and the exact results.  For the  subthreshold case our formulas do not agree with \cite{Andersson98}, but they do agree with the numerical results. This is shown in Fig. (\ref{subfig}).

\subsection{Fokker-Planck approximation}
The Fokker-Planck approach 
approximates the finite difference equation (\ref{taueq}) 
by the differential equation (\ref{FPE}), solves it, and then extracts the 
large $N$ asymptotics. This has the advantage of producing a tractable and generally 
useful partial differential equation, Eq. (\ref{FPEpde}). However, as we will see, it does not 
give the correct answers in general for our class of processes. 
In this section we will only be concerned with the superthreshold case.

\subsubsection{Asymptotic estimates}
The solution to Eq. (\ref{FPE}) is (neglecting ${\cal O}(1/N^2)$):
\begin{equation}
T(x) = \frac{1}{\myu{r}} \int_0^x e^{-N(V(r)-V(y))} \ dy \ + 
2N \int_0^x \int_y^r
\frac{e^{-N(V(z)-V(y))}}{\lam{z}+\myu{z}} \ dz \ dy
\end{equation}
where 
\begin{equation}
V(x) = -2 \int_0^x \frac{\lam{\xi}-\myu{\xi}}{\lam{\xi}+\myu{\xi}}\,d\xi \equiv
-2\int_0^x \frac{f(\xi)}{g^2(\xi)}d\xi
\label{veff}
\end{equation} 
plays the role of the effective potential.

\begin{figure}
\centerline{\includegraphics[width=4in]{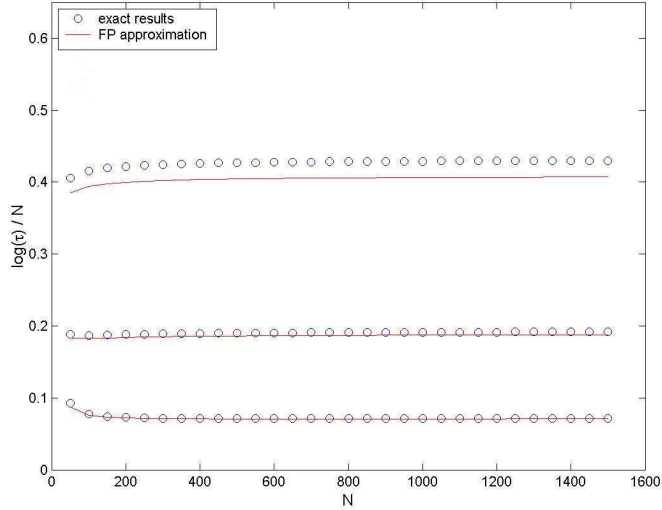}}
\caption{Comparison of FPE  and numerical results for the SIS model in the superthreshold case. 
$\log\taau{x_e}/N$ is plotted as a function of $N$ for several values of $\Lambda$. }
\label{FPfig}
\end{figure}

The asymptotic behavior, using standard techniques \cite{Gardiner83} is:
\begin{equation}
T(x)\approx T(x_e)=\frac{2}{|V'(0)|}\sqrt{\frac{2\pi}{NV''(x_e)}}
\frac{e^{-NV(x_e)}}{g^2(x_e)}\times \left( 1+{\cal
O}(1/N) \right).
\label{FPest}
\end{equation}
For the special case of the SIS model we find:
\begin{equation}
T(x)\approx
\frac{\Lambda+1}{2\sqrt{\Lambda}}
\frac{\Lambda}{(\Lambda-1)^2}\sqrt{\frac{2\pi}{N}} e^{-NV(x_e)}
\times \left( 1+{\cal O}(1/N) \right),
\label{FPSIS}
\end{equation}
where
\begin{equation}
-V(x_e)=\frac{4}{\Lambda}\log{\frac{2}{\Lambda+1}} +
2\frac{\Lambda-1}{\Lambda}.
\end{equation}
This expression is quite different from the first line of the table above (which agrees with the exact results). In Fig. (\ref{FPfig}) we show a comparison of Eq. (\ref{FPSIS}) with the exact results. It is clear that there is a significant discrepancy for large values of  $\Lambda$, that is, far from the threshold at $\Lambda=1$. Note that $\log\taau{x_e}/N$ is plotted in the figure. From the figure we see that, for $N=1000$, $T$ is a factor of about $10^7$ smaller than the exact result.

We have done a similar calculation for the logistic model.  The numerical results of Ref. \cite{Grasman97} are all near to threshold, so that the apparent numerical verification of their FPE calculation is only of limited validity. Far above threshold there are similar very large discrepancies between $T$ and $\bar\tau$.

\subsubsection{The effective potentials}
We now turn to the source of the problem with the FPE estimate of
$\bar\tau$. For the superthreshold case $\bar\tau$ in Eq. (\ref{FPest})
is of the standard form of a relatively slowly varying prefactor
multiplied by $\exp(-NV)$ where $V$ is given in Eq. (\ref{veff}). In the
following discussion we will ignore the prefactor and consider only the
dominant exponential term. The  FPE gives a $V$ which is not the same as
the correct answer, $\Phi$, Eq. (\ref{Phi}).

However, $V$ is close to  $\Phi$ quite near the threshold, in which case force is small over
the important range of $x$, namely $[0,x_e]$.
To see this, put  $\lam{x} - \myu{x}= f_N(x)$. Assume that the force
is, in fact, small: $f_N(x) \rightarrow 0$ 
appropriately uniformly in $x$ as $N \rightarrow \infty$. Then:
\begin{equation}
d\Phi(x)/dx = -\frac{f_N(x)}{\myu{x}} 
+ \frac{1}{2} \left( \frac{f_N(x)}{\myu{x}} \right)^2 
- \frac{1}{3} \left( \frac{f_N(x)}{\myu{x}} \right)^3
+ \dots
\end{equation} 
while
\begin{equation}
dV(x)/dx = -\frac{f_N(x)}{\myu{x}}
+ \frac{1}{2} \left(\frac{ f_N(x)}{\myu{x}} \right)^2
- \frac{1}{4} \left(\frac{f_N(x)}{\myu{x}} \right)^3
+ \dots .
\end{equation}
Hence the exponential terms formally agree to a factor of 
$(1+{\cal O}(1/N))$ only if 
$f_N(x) = \lam{x}-\myu{x} \le {\cal O}(N^{-1/3})$, that is, the drift must be small over the whole range of $x$. Equivalently, we can write  $\rho_n = \lambda_n/\mu_n = 1+ {\cal O}(1/N^{1/3+\epsilon}), \epsilon>0$, 
or, $x_e = {\cal O}(N^{-1/3-\epsilon})$. 
If this is true then the two estimates of $\bar\tau$ differ by factors of order unity.

\subsubsection{'Corrected' Fokker-Planck equation}
We might be tempted to define a `corrected'  FPE, with an effective potential $\Phi$ by redefining $f$ and $g^2$ in  Eq. (\ref{FPEpde}). A suitable $f$ and $g^2$ could be found to do this, but the corrected FPE would not be particularly useful. The point of using Eq. (\ref{FPEpde}) is to have a unified description of the process, equivalent in the continuum limit, to the original master equation. In particular, we should be able to describe the quasi-stationary distribution with the same equation. We will show that this is not possible.

A version of the quasi-stationary distribution \cite{Nasell99} is obtained by changing the boundary condition at the origin to reflecting, that is, we set $\lambda_0=1$. Then a stationary distribution exists. We can find this by returning to Eq. (\ref{master}), setting $d\pi_n(t)=0$, and solving the equation. The result is:
\begin{equation}
\pi_n=\frac{\prod_{j=0}^{n-1}[\lambda_j/\mu_{j+1}]}
{ 1+\sum_{k=1}^{\infty}\prod_{j=0}^{k-1}[\lambda_j/\mu_{j+1}]}
\label{steady}
\end{equation}
We take the continuum limit by defining $p(n/N)=N\pi_n$, and using the method of Eq. (\ref{trap}). We have:
\begin{equation}
p(x) \propto \sqrt{\frac{\lam{0} \myu{0}}{\lam{x} \myu{x}}} e^{-N\Phi(x)}(1+\cal{O}(1/N))
\label{steadycont}
\end{equation}

We can also solve for the stationary state of Eq. (\ref{FPEpde}), with the requirement that the effective potential be $\Phi$. This gives:
\begin{equation}
P(x) \propto e^{-N\Phi(x)}/g^2
\label{FPEsteady}
\end{equation}
Comparing these two equations we see that we must set:
\begin{eqnarray}
g^2(x) &=& C\sqrt{\lam{x}\myu{x}}  \nonumber \\
f(x) &=& C \sqrt{\lam{x}\myu{x}} \log \frac{\lam{x}}{\myu{x}},
\end{eqnarray}
where $C$ is a constant.
We can also recalculate $T$, as in Eq. (\ref{FPest}). If we compare to Eq. (\ref{super}) we see that while we have the correct effective potential (by construction), we do not get the correct prefactor, so that $T$ is inconsistent with $\bar\tau$.

\section{Summary and Discussion}
We have described two sorts of results in this paper. On the one hand, we have shown how to generalize previous work on the SIS model \cite{Weiss71,Oppenheim77,Nasell99,Andersson98}  to general birth-death processes, and to the threshold and subthreshold cases. We think that we have corrected an error in Ref. \cite{Andersson98} for the subthreshold case. 

Our treatment of the FPE should be viewed mainly as a warning about the subtlety of the relationship between discrete and continuum approaches. In cases where the continuum equation, Eq. (\ref{FPEpde}) `should'  work, it fails if the drift is too large. This is reflected in the fact  that the exact effective potential, $\Phi = -\int [\log(1+f/g^2)-\log(1-f/g^2)]$, is different from the FPE expression, $V=-2\int[f/g^2]$. They are the same for small $f$.
For practical purposes, if the process in not immediately above threshold, it is preferable to use Eq. (\ref{super}) or even the exact expression, Eq. (\ref{exact}). 

However, for a problem with a wide separation of scales, this option is often not available. For example, in work on modeling calcium waves in cells \cite{Jung03}, the underlying processes are often too complex to allow a practical exact calculation. One method to circumvent this difficulty is to use a Langevin equation, replacing some of the rapid processes by noise terms. In practice, however, there is some evidence that such a method gives acceptable results only near an appropriately defined threshold \cite{Jung03a}. 

We have tried to find a heuristic argument for the failure of the FPE for the extinction time. We speculate that in the large $N$ limit, the corresponding stochastic process will have discontinuous sample paths  since the fluctuations that lead to extinction probably occur over times that do not scale with $N$. In this case, no diffusion approximation  is possible.

\paragraph{Acknowledgements:}
This work was supported in part by NSF Award DMS-0244419.

\pagebreak
\appendix
{\bf \huge Appendix}

In this appendix we give the details of the large $N$ expansion of Eq. (\ref{exact}) which we repeat here for convenience:
\begin{equation}
\tau_n = \sum_{m=1}^{n} 
\big[ 
\frac{1}{\mu_m} +  \overbrace{ \prod_{i=1}^{m-1} \frac{1}{\rho_i}  
\ \sum_{j=m+1}^{R} \frac{1}{\mu_j} \ \prod_{k=1}^{j-1} \rho_k  }^{A_m}
\big].
\label{exact1}
\end{equation}
In the following sections we discuss the superthreshold, threshold, and subthreshold cases. The major computations are of the term $\sum A_m$ in Eq. (\ref{exact1}).

\section{ Superthreshold}


In this situation $\Phi(x)$ is convex with a quadratic minimum at $x_e$, 
the unique solution of $\lam{x_e}=\myu{x_e}$.
Thus, invoking standard integral estimates,
\begin{eqnarray}
\sum_{j=m+1}^{R}\frac{1}{\mu_j}\prod_{k=1}^{j-1}\rho_k &=& 
\sum_{j=m+1}^{R}\frac{1}{\mu_j}\frac{1}{\sqrt{\ro{0}\ro{j/N}}} 
e^{-N\Phi(j/N)} \times (1+{\cal O}(1/N)) \\ \nonumber
&=& \int_s^r \frac{dz}{\sqrt{\ro{0}\ro{z}}\myu{z}} e^{-N\Phi(z)} 
\times (1+{\cal O}(1/N)).
\end{eqnarray}
For the next step we use the fact that if $h$ is a smooth
function with $h(x_e)\neq 0$, then for $s < x_e$,
\begin{equation}
\int_s^r h(\xi)e^{-N\Phi(\xi)}d\xi = 
h(x_e) \ \sqrt{\frac{2\pi}{N\Phi^{''}(x_e)}} \
e^{-N\Phi(x_e)}\ \times \ \big(1+{\cal O}(1/N)\big)
\end{equation}
Hence, recalling $s=m/N$, and using $n=O(N)$, $\Phi^{'}(0)<0$, $\Phi(0)=0$ 
and the smoothness of $\bar{\rho}$ to compute the geometric sum after 
expanding $\Phi(m/N)$ around $0$,
\begin{eqnarray}
\sum_{m=1}^{n}A_m 
&=& \sum_{m=1}^{n}\frac{1}{\myu{x_e}} \sqrt{\frac{\ro{m/N}}{\ro{x_e}}} 
\sqrt{\frac{2\pi}{N\Phi^{''}(x_e)}} e^{-N\Phi(x_e)} e^{N\Phi(m/N)} 
 \times \big(1+{\cal O}(1/N)\big)  \nonumber \\
&=& \sqrt{\frac{2\pi}{N\Phi^{''}(x_e)}} \
\frac{e^{-N\Phi(x_e)}}{\myu{x_e}\sqrt{\ro{x_e}}} \ 
\sum_{m=1}^{n} \sqrt{\ro{m/N}} e^{N\Phi(m/N)} 
\times \big(1+{\cal O}(1/N)\big)  \nonumber \\ 
&=& \sqrt{\frac{2\pi}{N\Phi^{''}(x_e)}} \
\frac{e^{-N\Phi(x_e)}}{\myu{x_e}\sqrt{\ro{x_e}}} \
\frac{\sqrt{\ro{0}}}{1-e^{\Phi^{'}(0)}}
\times \big(1+{\cal O}(1/N)\big).
\label{last}
\end{eqnarray}
The last expression is exponentially large in $N$ so 
$\sum_{m=1}^{n}\frac{1}{\mu_m}$, which we estimate in the subthreshold 
case, below, is completely negligible.
Eliminating $\bar{\rho}$, then, we find
\begin{equation}
\tau_n = \sqrt{\frac{2 \pi \lam{0}\myu{0}}{N 
[\lam{x_e}{\bar{\mu}}^{'}(x_e)-{\bar{\lambda}}^{'}(x_e)\myu{x_e}]}} \ 
\frac{e^{-N\Phi(x_e)}}{\lam{0}-\myu{0}} 
\times \big(1+{\cal O}(1/N)\big),
\end{equation}
independent of $n$ for $n={\cal O}(N)$.
The boundary layer correction follows simply from a finite geometric 
series approximation to the sum in the penultimate line in (\ref{last}).

\section{Threshold}
Here the potential function $\Phi(x)$ is convex with vanishing derivative 
but nonvanishing curvature at $x = 0$.
Referring to the terms in the exact solution (\ref{exact1}), invoking 
the trapezoid approximation, and putting $s=m/N$,
\begin{eqnarray}
\sum_{m=1}^n A_m &=& \sum_{m=1}^n \int_{m/N}^r
\frac{\sqrt{\ro{m/N}}}{\sqrt{\ro{z}}\myu{z}} 
\ \ e^{-N(\Phi(z)-\Phi(m/N))}dz \ \times \ \big(1+{\cal O}(1/N)\big)
\nonumber \\
&=& N\int_0^x ds \int_s^r dz \ \frac{\sqrt{\ro{s}}}{\sqrt{\lam{z}\myu{z}}} 
\ e^{-N(\Phi(z)-\Phi(s))}\ \times \ \big(1+{\cal O}(1/N)\big).
\end{eqnarray}
Most of the contribution to the integral comes from a neighborhood of 
the origin, so we proceed making standard integral approximations.
Expand $\Phi(x)$, $\ro{s}$, $\lam{z}$ and $\myu{z}$ using 
$\lam{0}=\myu{0}=\Phi(0)=\Phi^{'}(0)=0$:
\begin{equation}
\sum_{m=1}^n A_m = 
N\frac{\sqrt{\ro{0}}}{\sqrt{\bar{\lambda}^{'}(0)\bar{\mu}^{'}(0)}} 
\int_0^x ds \int_s^r dz \ \frac{e^{-N\Phi^{''}(0)(\frac{z^2}{2} 
-\frac{s^2}{2})}}{z} \ \times \ \big(1+{\cal O}(1/N)\big).
\end{equation}
Switching the integrals,
\begin{equation}
\sum_{m=1}^n A_m =
N \frac{\sqrt{\ro{0}}}{\sqrt{\bar{\lambda}^{'}(0)\bar{\mu}^{'}(0)}} 
\int_0^r \frac{e^{-N\Phi^{''}(0)z^2/2}}{z}dz \int_0^{\text{min}(z,x)} 
e^{N\Phi^{''}(0)s^2/2}ds \ \times \ \big(1+{\cal O}(1/N)\big),
\end{equation}
and changing variables,
\begin{equation}
= \sqrt{\frac{N}{\Phi^{''}(0)}} 
\frac{\sqrt{\ro{0}}}{\sqrt{\bar{\lambda}^{'}(0)\bar{\mu}^{'}(0)}} 
\int_0^{r\sqrt{N\Phi^{''}(0)}}\frac{e^{-v^2/2}}{v}dv 
\int_0^{\text{min}(v,\sqrt{N\Phi^{''}(0)}x)}e^{u^2/2}du 
 \ \times \ \big(1+{\cal O}(1/N)\big).
\end{equation}
Then because $N$ is large we may modify the limits of the integrals.
\begin{equation}
\sum_{m=1}^n A_m = 
\sqrt{\frac{N}{\Phi^{''}(0)}} \ 
\frac{\sqrt{\ro{0}}}{\sqrt{\bar{\lambda}^{'}(0)\bar{\mu}^{'}(0)}} \ 
\int_0^{\infty} dv \frac{e^{-v^2/2}}{v} 
\int_0^v du e^{u^2/2} \times \big(1+O(1/N)\big).
\end{equation}
Integrate $e^{u^2/2}$ using its Taylor expansion and use the formula 
for the even moments of a gaussian to obtain
\begin{eqnarray}
\sum_{m=1}^n A_m 
&=& \sqrt{\frac{N}{\Phi^{''}(0)}} 
\frac{\sqrt{\ro{0}}}{\sqrt{\bar{\lambda}^{'}(0)\bar{\mu}^{'}(0)}} 
\sum_{k=0}^{\infty}\frac{1}{2^k k! (2k+1)}
\int_0^{\infty}v^{2k} e^{-v^2/2}dv\big(1+{\cal O}(1/N)\big) \\ \nonumber
&=& \sqrt{\frac{N}{\Phi^{''}(0)}} \
\frac{\sqrt{\ro{0}}}{\sqrt{\bar{\lambda}^{'}(0)\bar{\mu}^{'}(0)}} \
\sum_{k=0}^{\infty}\frac{(2k)!}{(2^k k!)^2(2k+1)} 
\sqrt{\frac{\pi}{2}} \times \big(1+{\cal O}(1/N)\big).
\end{eqnarray}
The sum above converges to $C \approx 1.5687$, so
\begin{equation}
\sum_{m=1}^n A_m = C\ \sqrt{\frac{\pi}{2\Phi^{''}(0)}} \
\sqrt{ \frac{\ro{0}}{\bar{\lambda}^{'}(0)\bar{\mu}^{'}(0)} } \
\sqrt{N} \ + \ {\cal O}(\frac{1}{\sqrt{N}})
\end{equation}
We show in the next section that
\begin{equation}
\sum_{m=1}^{n}\frac{1}{\mu_m} \ = \ \frac{1}{\mb} \log{n} + {\cal O}(1).
\end{equation}
Thus,
\begin{equation}
\tau_n \ = \ C \ \sqrt{\frac{\pi}{2\Phi^{''}(0)}} 
\sqrt{\frac{\ro{0}}{\bar{\lambda}^{'}(0)\bar{\mu}^{'}(0)}} 
\ \sqrt{N} \ + \ \frac{1}{\mb}\log{n} + {\cal O}(1) 
\end{equation}

\section{Subthreshold case}
Here the potential function $\Phi(x)$ is a convex function with positive 
derivative at 0.
From (\ref{exact1}),
\begin{eqnarray}
A_m &=& \sqrt{\ro{m/N}}\sum_{j=m+1}^{R} 
\frac{1}{N\myu{j/N}\sqrt{\ro{j/N}}} e^{N(\Phi(m/N)-\Phi(j/N))} 
\big(1+{\cal O}(1/N)\big) \\ \nonumber
&=& \sqrt{\ro{m/N}}\sum_{j=m+1}^{R} 
\frac{\ro{m/N}^{j-m}}{N\sqrt{\lam{j/N}\myu{j/N}}} 
\big(1+{\cal O}(1/N)\big) \\ \nonumber 
&=&\ro{m/N}^{-m+1/2}\sum_{j=m+1}^{R} 
\frac{\ro{m/N}^j}{N\sqrt{\lam{j/N}\myu{j/N}}} \big(1+{\cal O}(1/N)\big)
\end{eqnarray}
Adding and subtracting 
\begin{equation}
\sum_{j=m+1}^{R} \frac{\ro{m/N}^{j-m+1/2}}{N\sqrt{\lam{m/N}\myu{m/N}}}
\end{equation}
and recalling that in subthreshold case $\ro{x}<1$ uniformly for
$x\in[0,r]$, we have
\begin{eqnarray}
A_m &=& \ro{m/N}^{-m+1/2}\sum_{j=m+1}^{R}\big( 
\frac{\ro{m/N}^j}{N\sqrt{\lam{j/N}\myu{j/N}}} - 
\frac{\ro{m/N}^j}{N\sqrt{\lam{m/N}\myu{m/N}}}\big) \\ \nonumber
&+& \sum_{j=m+1}^{R}\frac{\ro{m/N}^{j-m+1/2}}{N\sqrt{\lam{m/N}\myu{m/N}}} 
\\ \nonumber
&=& \underbrace{\ro{m/N}^{-m+1/2} 
\int_{s}^{r}\big(\frac{\ro{s}^{N\xi}}{\sqrt{\lam{\xi}\myu{\xi}}} 
-\frac{\ro{s}^{N\xi}}{\sqrt{\lam{s}\myu{s}}}\big)d\xi}_{B_m} \\ \nonumber
&+& \underbrace{\frac{\ro{m/N}}{1-\ro{m/N}}\frac{1}{N\myu{m/N}}}_{C_m} 
+ {\cal O}(1/N) \\ \nonumber
&=& B_m \ + \ C_m \ + \ {\cal O}(1/N).
\end{eqnarray}

Now we evaluate $\sum_{m=1}^{n}B_m$ and $\sum_{m=1}^{n}C_m$ separately as 
$n=Nx\rightarrow\infty$.
Changing variables in the integral in $B_m$ and using the facts that
for small $s$, $\sqrt{\lam{s}\myu{s}} = \sqrt{\lam{0}\myu{0}} \ s + {\cal 
O}(s^2)$, or $\sqrt{\lam{s}\myu{s}}=\sqrt{\lam{0}\myu{0}} \ s \times 
(1+{\cal O}(s))$, and $\ro{s}=O(1)$, we find
\begin{eqnarray}
B_m &=&
\int_{0}^{r-s}\big(\frac{\ro{s}^{N\xi+1/2}}{\sqrt{\lam{\xi+s}\myu{\xi+s}}} 
-\frac{\ro{s}^{N\xi+1/2}}{\sqrt{\lam{s}\myu{s}}}\big)d\xi \\ \nonumber 
&=& {\cal O}(1) \times \int_{0}^{r-s}\ro{s}^{N\xi+1/2} 
\frac{\xi}{s(\xi+s)}d\xi\  \\ \nonumber
&=& {\cal O}(1) \times \int_{0}^{r-s} 
e^{N\xi\log{\ro{s}}}\frac{\xi}{s(\xi+s)}d\xi \\ \nonumber
&=& {\cal O}(1) \times \int_{0}^{r/s-1}e^{(s\alpha)z}\frac{z}{1+z}dz  \\ 
\nonumber
&\leqq& {\cal O}(1) \times \int_{0}^{r/s-1}e^{(s\alpha)z}zdz  \\ \nonumber 
&\leqq& {\cal O}(1) \times \frac{1}{{(s\alpha)}^2}
\end{eqnarray}
where we denote $\alpha=N \log{\ro{s}}$.
Then recalling $s=m/N$, we conclude
\begin{equation}
\sum_{m=1}^{n}B_m \ \leqq \
{\cal O}(1) \times \sum_{m=1}^{n} 
\frac{1}{m^2{(\log{\ro{m/N}})}^2} \ = \ {\cal O}(1)
\end{equation}
as $N\rightarrow\infty$ (which means $n\rightarrow\infty$ as well).

On the other hand the $C_m$ terms may be written
\begin{equation}
\sum_{m=1}^{n}C_m = \sum_{m=1}^{n} \eta(\frac{m}{N}) \frac{1}{N\myu{m/N}}
\end{equation}
where we define 
\begin{equation}
\eta(\frac{m}{N}) = \frac{\ro{m/N}}{1-\ro{m/N}}.
\end{equation}
Then subtract and add the `divergent part' of the sum:
\begin{eqnarray}
\sum_{m=1}^{n}C_m &=&  
\sum_{m=1}^{n}\big[\frac{\eta(\frac{m}{N})}{N\myu{m/N}} 
-\frac{\eta(0)}{m\mb}\big]+\sum_{m=1}^{n}\frac{\eta(0)}{m\mb} \\ \nonumber
&=& \underbrace{\int_0^x \frac{\eta(\xi)\xi\mb - 
\eta(0)\myu{\xi}}{\xi\mb\myu{\xi}}d\xi}_{{\cal O}(1)} \ + \
\frac{\eta(0)}{\mb}(\gamma+\log{n})\ 
+ \ {\cal O}(\frac{1}{N}) \\ \nonumber
&=& \frac{\eta(0)}{\mb} \log{(n)} \ + \ {\cal O}(1)
\end{eqnarray}
where $\gamma = 0.5772..$ is Euler's constant.

Finally,
\begin{eqnarray} 
\sum_{m=1}^{n}\frac{1}{\mu_m} &=& 
\sum_{m=1}^{n}\frac{1}{N\myu{m/N}} \\ \nonumber
&=& \frac{\gamma+\log{(n)}}{\mb} 
+ \underbrace{\int_{0}^{x}\frac{\xi\mb-\myu{\xi}} 
{\xi\mb\myu{\xi}}d\xi}_{O(1)} \ + \ {\cal O}(1/N) \\ \nonumber
&=& \frac{1}{\mb}\log{(n)} + {\cal O}(1).
\end{eqnarray}
Putting these calculations together we conclude
\begin{equation}
\tau_n \ = \ \taau{x} \ = \ \frac{\eta(0)}{\mb}\log{n} + 
\frac{1}{\mb}\log{n} 
+ {\cal O}(1) \ = \ \frac{1}{\mb(1-\ro{0})}\log{Nx} + {\cal O}(1).
\end{equation}

\bibliography{mfpt1}

\end{document}